\documentclass[sigconf]{acmart}
\usepackage{multicol, multirow}
\usepackage{subfigure}
\usepackage{url}
\usepackage{makecell}
\AtBeginDocument{%
  }

\setcopyright{acmlicensed}
\copyrightyear{2018}
\acmYear{2018}
\acmDOI{XXXXXXX.XXXXXXX}
\acmConference[Conference acronym 'XX]{Make sure to enter the correct
  conference title from your rights confirmation email}{June 03--05,
  2018}{Woodstock, NY}
\acmISBN{978-1-4503-XXXX-X/2018/06}

\begin{document}

\title{A Novel Mamba-based Sequential Recommendation Method}

\author{Jun Yuan}
\email{yuanjun25@huawei.com}
\affiliation{%
  \institution{Huawei Technologies Co., Ltd.}
  \city{Shenzhen}
  \state{Guangdong}
  \country{China}
}

\renewcommand{\shortauthors}{Jun Yuan}


\begin{abstract}
  Sequential recommendation (SR), which encodes user activity to predict the next action, has emerged as a widely adopted strategy in developing commercial personalized recommendation systems. Although Transformer-based models have proven effective for sequential recommendation, the complexity of the self-attention module in Transformers scales quadratically with the sequence length. Controlling model complexity is essential for large-scale recommendation systems, as these systems may need to handle billion-scale vocabularies that evolve continuously, as well as user behavior sequences that can exceed tens of thousands in length. In this paper, we propose a novel multi-head latent Mamba architecture, which employs multiple low-dimensional Mamba layers and fully connected layers coupled with positional encoding to simultaneously capture historical and item information within each latent subspace. We evocatively name our method \textit{Hydra}, after the many-headed serpent in Greek mythology. Our proposed Hydra not only enables scaling up to large-scale parameters but also extends to multi-domain recommendation by integrating and fine-tuning LLMs. Through extensive experiments on public datasets, we demonstrate how Hydra effectively addresses the effectiveness-efficiency dilemma, outperforming state-of-the-art sequential recommendation baselines with significantly fewer parameters and reduced training time.
\end{abstract}


\maketitle

\section{Introduction}
Recommendation systems play a pivotal role in current online content platforms and e-commerce. The recommendation algorithm is a complex problem that requires extracting user interests to predict future behaviors across billions of items. Nowadays, sequential recommendation (SR) has occupied a dominant position in commercial recommendation systems, including e-commerce~\cite{chen2019behaviorsequencetransformerecommerce}, social media~\cite{xia2023transacttransformerbasedrealtimeuser}, news/video feeds~\cite{10.1145/3477495.3531862}, and online advertising~\cite{10.1145/3219819.3219823}. The goal of sequential recommendation systems is to combine personalized models of user behavior (based on historical activities) with a notion of "context" derived from users’ recent actions~\cite{kang2018selfattentivesequentialrecommendation}. Sequential recommendation has been explored for years, and various SR models have been proposed~\cite{10.1145/2766462.2767694,bert4rec,10.1609/aaai.v33i01.33015941}. 

In recent years, the remarkable success of large language models (LLMs)~\cite{yang2024qwen2technicalreport,touvron2023llamaopenefficientfoundation,openai2024gpt4technicalreport}, exemplified by GPT~\cite{openai2024gpt4technicalreport}, has prompted many studies to introduce relevant experiences into the field of sequential recommendation~\cite{Wu2023ASO}. The primary research and explorations can be categorized into two main types:

One line of studies has made significant efforts to apply \textit{Transformers} or variant to sequential recommendation, yielding many notable achievements, such as SASRec~\cite{kang2018selfattentivesequentialrecommendation}, BST~\cite{chen2019behaviorsequencetransformerecommerce} and others~\cite{10.1145/3340531.3411954,10.1145/3581783.3611967}. Some researches~\cite{zhai2024hstu,shen2025optimizingsequentialrecommendationmodels} further demonstrates that there may exist a certain "scaling law" in recommendation, where larger networks and increased data volumes may lead to better recommendation performance. However, two essential differences distinguish sequential recommendation from generative language models: \underline{First}, user interaction sequences in sequential recommendation systems are often much longer than token sequences in language, potentially reaching hundreds of thousands of interactions~\cite{zhai2024hstu}. Moreover, the relationships between items are not as tight as those between tokens in sentences, with many items possibly being noise. \underline{Second}, items in user interaction sequences rarely exhibit the strict logical relationships found in language; instead, they are primarily characterized by collaborative relationships~\cite{10.1145/3627673.3679558}. In addition to these two differences, online recommendation systems necessitate a stricter adherence to processing latency constraints than current LLMs. Consequently, applying Transformers in sequential recommendation faces significant challenges.

Another line of studies~\cite{10.1145/3627673.3679558,10.1145/3640457.3688104} attempts to harness the "world knowledge" embedded in LLMs to enhance existing recommendation models, such as LEARN~\cite{jia2024learnknowledgeadaptationlarge} and HLLM~\cite{chen2024hllmenhancingsequentialrecommendations}. These researches have shown the significant effects to integrate LLMs and SR models, particularly in cold start scenarios or small domains. However, the rich semantic information contained in LLMs may conflict with the collaborative information inherent within recommendation data. Therefore, it is often necessary to fine-tune LLMs using recommendation data to achieve better performance in large domains~\cite{zhang2024collmintegratingcollaborativeembeddings,zhang2024textlikeencodingcollaborativeinformation}. These approaches necessity causes a potential issue: when a recommendation system encompasses multiple domains, the cost of fine-tuning an LLM for each domain will become prohibitive. Actually, the effect of LLMs in the multi-domain recommendation field is underexplored.

To address the above problems in existing work, this paper proposes a multi-head latent Mamba architecture to enhance sequential recommendation, based on Mamba-2~\cite{dao2024mamba2} architecture. Our proposed method first maps the item representation into multiple latent subspaces. It then utilizes multiple low-dimensional Mamba layers and fully connected (FC) layers coupled with positional encoding to simultaneously capture two distinct yet interrelated information of item in user context within each subspace: the \textbf{historical information} and \textbf{item information}. Historical information is derived from the entire user context before present user action, and the item information is extracted from item representations themselves. Subsequently, our method interacts and merges the multi-view historical and item information, enabling the efficient and effective processing of long and noisy user behavior sequences. Since our proposed method interweaves multiple Mambas to form a multi-head network, we evocatively name our method \textbf{Hydra}, after the many-headed serpent in Greek mythology.

Furthermore, extensive experiments demonstrate that Hydra not only enables the network to scale up, thereby effectively modeling large-scale recommendation scenarios, but also can be extended to multi-domain recommendation by integrating and fine-tuning LLMs. Experiments show that fine-tuning a single LLM can improve recommendation performance across various recommendation domains, significantly reducing the cost of fine-tuning.

The major contributions of this paper are as follows. 

\begin{enumerate}
    \item We first propose a novel multi-head latent Mamba architecture. Based on the architecture, an efficient sequential recommendation model, named Hydra, has been proposed. This approach enables efficient and effective interaction between historical and item information extracted from long and noisy user contexts. Moreover, Hydra can effectively transfers the world knowledge embedded in LLMs into the recommendation model.
    \item Hydra can improve multi-domain recommendation by integrating and fine-tuning LLMs on multi-domain recommendation data. To the best of our knowledge, Hydra first demonstrates that fine-tuning a single LLM can enhance recommendation performance across various domains.
    \item Extensive experiments demonstrate that Hydra significantly outperforms state-of-the-art (SOTA) SR baselines on several large-scale benchmark datasets in both single- and multiple-domain recommendation. 
\end{enumerate}

\section{Related Work}

\subsection{Sequential Recommendation} Early sequential methods like Markov chains~\cite{10.1145/2766462.2767694} were proposed for modeling user’s sequential patterns. However, these methods only capture local sequential behaviors between adjacent items. As a result, RNN/CNN-based models~\cite{hidasi2016sessionbasedrecommendationsrecurrentneural,10.1145/3289600.3290975} were developed to model higher-order relationships. Following the significant advancements in NLP, various attention-based SR models have been proposed~\cite{10.1145/3219819.3219823,10.1609/aaai.v33i01.33015941,kang2018selfattentivesequentialrecommendation,chen2019behaviorsequencetransformerecommerce,10.1145/3340531.3411954,bert4rec}, which have become a critical component of modern commercial recommendation systems. Attention-based SR models can be categorized into two categories: target attention based models~\cite{10.1145/3219819.3219823,10.1609/aaai.v33i01.33015941} and self-attention based models~\cite{kang2018selfattentivesequentialrecommendation,chen2019behaviorsequencetransformerecommerce,10.1145/3340531.3411954,bert4rec}.
Long-range sequence processing poses a significant challenge for Transformers due to their quadratic complexity in input length. A promising alternative is Mamba. Mamba4Rec~\cite{liu2024mamba4rec} first introduce Mamba into SR by simply replace self-attention in Transformers by a Mamba layer. However, the length-generalization capabilities of Mamba are relatively limited~\cite{benkish2024decimambaexploringlengthextrapolation}. Our proposed Hydra tries to use multi-head Mamba to improve the capabilities to handle long and noisy user context length.

\subsection{Recommendation with Language Models}
Existing work predominantly utilizes LLMs to generate knowledge-rich texts or utilizes LLM-derived embeddings as features to improve SR. These explorations can be categorized into two types. 

Firstly, LLMs are used for summarizing or supplementing information about users or items~\cite{jia2024learnknowledgeadaptationlarge,10.1145/3589334.3645458,10.1145/3340531.3411954,10.1145/3580305.3599519,10.1145/3581783.3611967}. For instance, RLMRec~\cite{10.1145/3589334.3645458} develops a user/item profiling paradigm empowered by LLMs, and aligns the semantic space of LLMs with the representation space of collaborative relational signals through a cross-view alignment framework. Recformer~\cite{10.1145/3580305.3599519} learns both textual and sequential patterns for recommendations by proposing a framework including pre-training and fine-tuning. MISSRec~\cite{10.1145/3581783.3611967} proposes a novel multi-modal pre-training and transfer learning framework, effectively addressing the cold-start problem and enabling efficient domain adaptation. In this paper, we focus solely on the ID features, though CAPE can be easily extended to multi-modal features.

Secondly, some works that have adapted LLMs for recommendation tasks, allowing their inputs or outputs to go beyond just textual forms\cite{10.1145/3627673.3679558}. LLaRA~\cite{10.1145/3626772.3657690} proposed a novel hybrid prompting method that integrates ID-based item embeddings with textual item features. LEARN~\cite{jia2024learnknowledgeadaptationlarge} utilizes pre-trained LLMs to extract item features. KAR~\cite{10.1145/3640457.3688104} adopts LLMs to generate user preference reasoning and item factual knowledge, enhancing RSs through hybrid-expert adaptors. HLLM~\cite{chen2024hllmenhancingsequentialrecommendations} uses a LLM to encode item feature and another to extract user interests. However, these methods are all based on the \textit{Transformers}, meaning that model complexity scales quadratically with the sequence length.

\subsection{Generative Recommendation}
Generative Recommendation reformulates recommendation problems as sequential transduction tasks within a generative modeling framework. TIGER~\cite{rajput2023recommender} first proposed the concept of "generative recommendation" for domain-level zero-shot recommendation. HSTU~\cite{zhai2024hstu} is yet another attempt to adapt Transformers to generative recommendation, after the TIGER model. It uses a point-wise normalization mechanism instead of softmax normalization, making it suitable for non-stationary vocabularies in streaming settings. HLLM~\cite{chen2024hllmenhancingsequentialrecommendations} can also be seen as a generative recommendation model. Generative recommendation models can scale up billions parameters which can significantly improve recommendation performance with large-scale data.

\begin{figure*}[htp]
    \centering
    \subfigure{
        \includegraphics[width=0.3\textwidth,height=0.25\textwidth]{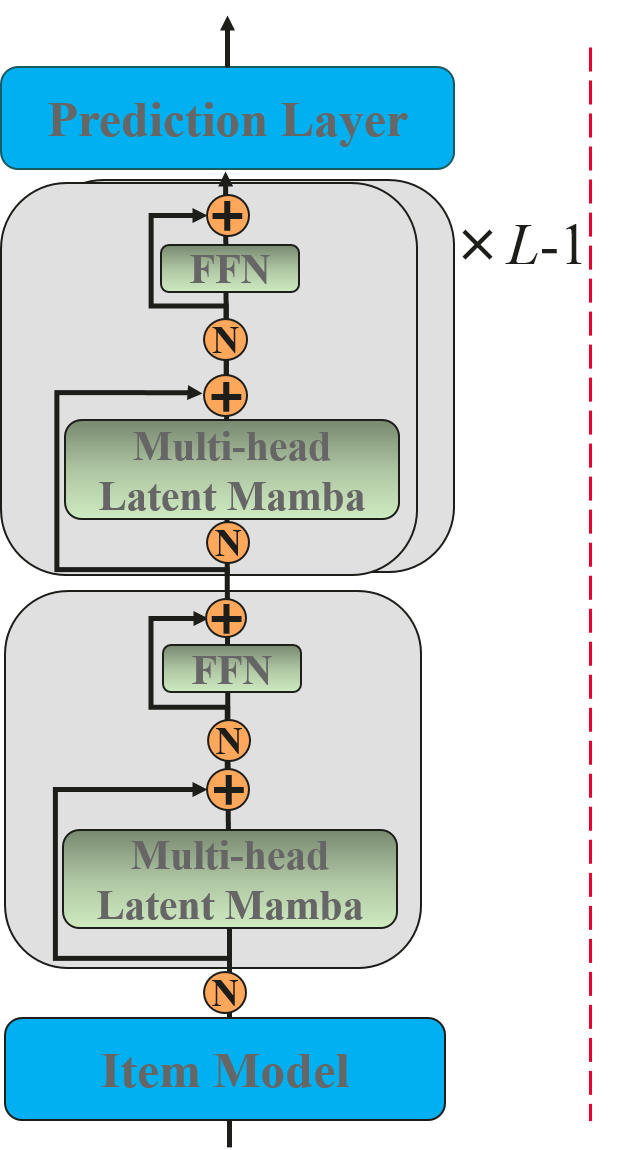}}\vspace{-3mm}
    \subfigure{\includegraphics[width=0.55\textwidth,height=0.25\textwidth]{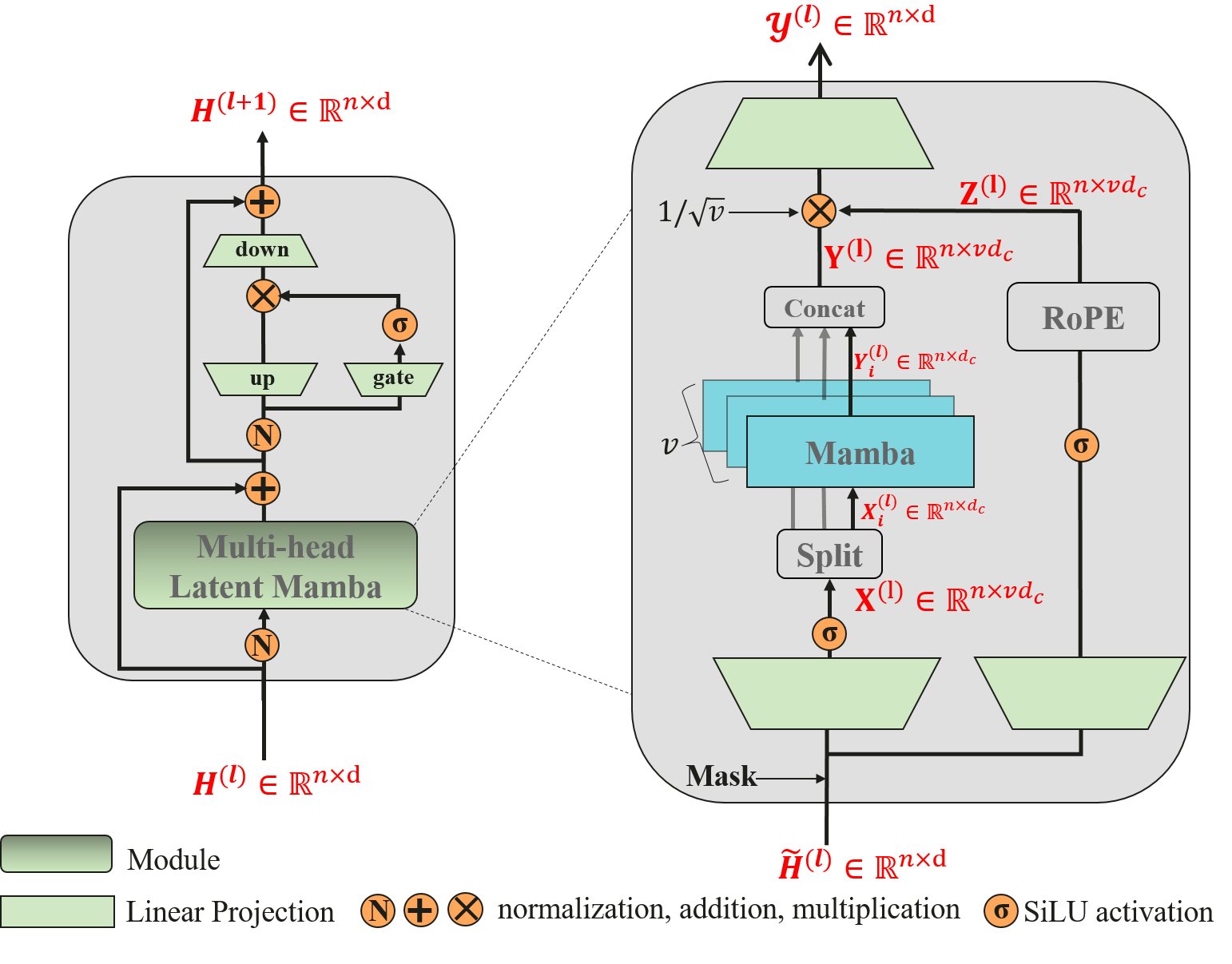}}\vspace{-3mm}
    \caption{\textit{Overall Architecture of Hydra}. Hydra has an item model, several identical stacked Hydra layers and a prediction layer. The core of Hydra layer is multi-head latent Mamba, which can extract and interact historical and item information to better model long and noisy user context.}\label{fig:arch}
    \vspace{-4mm}
\end{figure*}

\section{Problem Description}\label{sec:pre}
In this section, we first introduce some necessary background knowledge and then formally define the sequential recommendation problem. Through these discussions, we also introduce several important concepts and clearly define the associated notations.

\subsection{Preliminaries}
\textbf{Mamba}~\cite{dao2024mamba2} (or State Space Machine (SSM)) Given a 1D continuous input $x_{t} \in \mathbb{R}$, Mamba transforms it to $y_{t} \in \mathbb{R}$ via a learnable hidden state $H_{t} \in \mathbb{R}^{d}$ by introducing a selection mechanism that allows for input-dependent sequence processing This allows the model’s parameters to be adjusted dynamically according to the inputs and filter out irrelevant information. The SSM layer is based on a time-variant SSM with discretized parameters $\widebar{\boldsymbol{A}} \in \mathbb{R}^{d\times d}$, $\widebar{\boldsymbol{B}} \in \mathbb{R}^{1\times d}$ and $\widebar{\boldsymbol{C}} \in \mathbb{R}^{1\times d}$, which can be elaborated by the following recurrent rule:
\begin{equation}
\begin{aligned}
    &H^{'}_{t} = \widebar{\boldsymbol{A}}H_{t} + \widebar{\boldsymbol{B}}x_{t}\\
    &y_{t} = \widebar{\boldsymbol{C}}H_{t}
\end{aligned}
\end{equation} The time and complexity of Mamba is $O(d^{2})$. Therefore, reducing embedding dimension will reduce complexity quadratically. 

\textbf{Mamba-2}~\cite{dao2024mamba2} restricts the matrix $\boldsymbol{A}$ that all the diagonal elements are the same value. Additionally, it adopts different matrices for different SSM head to enable much larger state size. Mamba-2 is much more efficient than Mamba, and we propose a multi-head latent Mamba architecture based on Mamba-2, which can improve model capability and efficiency at the same time. Detailed complexity analysis please refer to Section~\ref{sec:complex}.


\subsection{Problem Definition}
We focus on the task of sequential recommendation, which is formally defined as follows: Given a user $u \in \mathcal{U}$ and a sequence of the user's historical interactions (referred to as the context) $U = \{I_1, I_2, \dots, I_n\}$ arranged in chronological order, the objective is to recommend the next most likely item $I_{n+1}$, where $n$ is the length of $U$ and $I \in \mathcal{I}$. $\mathcal{I}$ represents the set of all items, \emph{i.e.} $\forall I \in \mathcal{I}$ and $\mathcal{U}$ represents the set of all users in dataset.
\begin{equation}
   \max_{I_{n+1} \in \mathcal{I}} P(I_{n+1}\ |\ U)
\end{equation}
Each item $I$ is associated with an ID and additional features (e.g., category, tags, text, etc.). 

\begin{table}[htp]
\caption{Summary of notations}\vspace{-4mm}
\label{tab:notations}
\centering
\begin{tabular}{c|c}
\toprule
\textbf{Notation}&\textbf{Description}\\\hline
$I$ &Context item\\
$n$&Length of context\\
$d$&Dimension of input item embedding\\
$d_{c}$&Dimension of multi-head latent Mamba\\
$L$&Then number of network layers\\
$v$&The number of head in each layer\\
$\mathbf{H}^{(l)}$ &Input of \emph{l}-th layer, $\mathbb{R}^{n \times d}$, $l \in [1, L]$\\
$\mathbf{H}^{(l)}_{i}$ &Input of \emph{i}-th item of \emph{l}-th layer,$i \in [1, n]$, $\mathbb{R}^{d}$\\
\bottomrule
\end{tabular}
\vspace{-4mm}
\end{table}

\section{The Architecture of Hydra}

In this section, we introduce the overall network architecture of Hydra, which is depicted in Figure~\ref{fig:arch}. The training objectives will be discussed in Section~\ref{sec:obj}.

We begin with a high-level overview of the method, followed by an exploration of its technical components. As shown in the left part of Figure~\ref{fig:arch}, Hydra constructs a sequential recommendation model through an item model, several stacked Hydra layers, and a prediction layer. Subsequently, we describe the multi-domain extension of Hydra. Last but not least, we analyze the time and space complexity of our method, which demonstrate the superior efficiency of Hydra.

\subsection{Item Model}
Item model is designed to extract item features. As shown in Figure~\ref{fig:item_model}, Hydra can use two types of Item model depending on different situations: Embedding Layer and Item LLM. Besides, item model does not explicitly incorporate positional information, while positional encoding method will be integrated within the Hydra layers.

\subsubsection{Embedding Layer}
Consistent with existing models, our approach utilizes an embedding layer to map item IDs to a high-dimensional space (right part of Figure~\ref{fig:item_model}). The embedding layer uses a learnable embedding matrix $\mathbf{E} \in \mathbb{R}^{|\mathcal{I}|\times d}$, where $d$ is the embedding dimension. By applying the embedding layer to the input item sequence $U$, we obtain the initial item embeddings $\mathbf{H}^{(0)} \in \mathbb{R}^{n \times d}$. To enhance robustness and prevent overfitting, we incorporate both embedding dropout and RMSNorm~\cite{zhang2019rootmeansquarelayer} after retrieving the embeddings:
\begin{equation}
    \mathbf{H}^{(1)} = \mathrm{RMSNorm}(\mathrm{Dropout}(\mathbf{H}^{(0)}))
\end{equation}

\subsubsection{Item LLM}
Hydra can employ a LLM to map text description of an item to an embedding representation. Note that item LLM will be fine-tuned using recommendation data during training Hydra. 

\begin{figure}
    \centering
    \includegraphics[width=0.85\linewidth,height=0.28\textwidth]{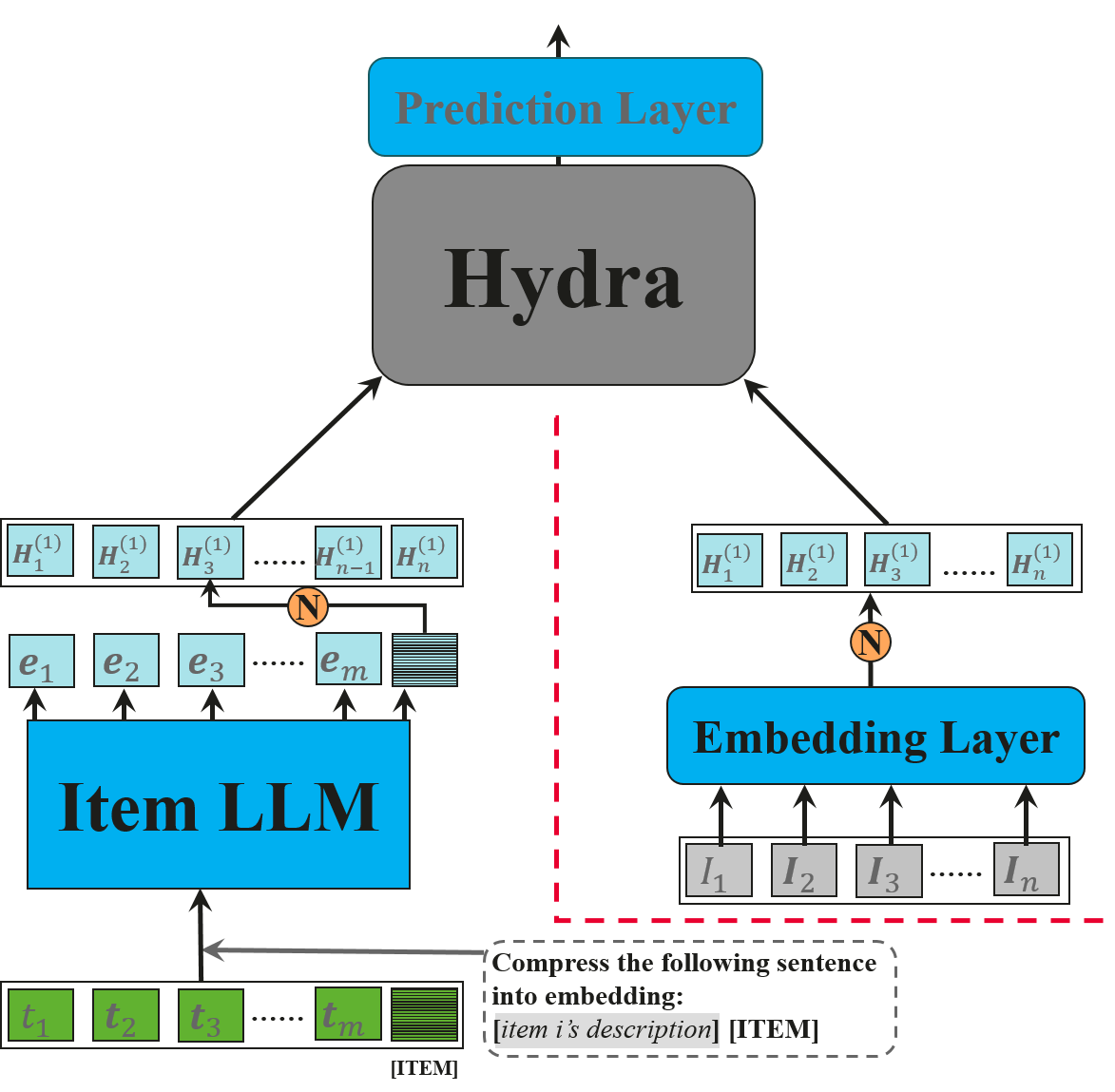}\vspace{-2mm}
    \caption{Two types of Item Model in Hydra. Embedding layer can represent ID features of item. Item LLM can be any pre-trained LLM and encode text into hidden embedding. The gray box marked Hydra is actually the stacked Hydra layers.}\vspace{-6mm}
    \label{fig:item_model}
\end{figure}

Inspired by previous researchers~\cite{chen2024hllmenhancingsequentialrecommendations}, as shown in the left part of Figure~\ref{fig:item_model}, for Item $I_{i}$, we first flatten its corresponding textual attributes into a sentence and prepend it with a fixed prompt. After passing through the LLM tokenizer, we additionally append a special token $[ITEM]$ at the end. The input token sequence for the Item LLM can be formulated as $\{t_1, t_2, . . . , t_m, [ITEM]\}$, where $m$ represents the length of text tokens. The hidden state corresponding to the special token $[ITEM]$ from the last layer is considered as the item embedding $\mathbf{H}_{i}^{(0)}$.
\begin{equation}
    \mathbf{H}_{i}^{(0)} = \mathrm{LLM}(\{t_1, t_2, . . . , t_m, [ITEM]\})_{m+1}
\end{equation} Similarly, to enhance robustness and prevent overfitting, we also apply both dropout and RMSNorm on item embedding as aforementioned manner.


\subsection{Hydra Layers}
As shown in right part of Figure~\ref{fig:arch}, the core of Hydra is the Hydra layer, which consists of a multi-head latent mamba module coupled with a feed-forward network.

\subsubsection{Multi-head Latent Mamba}
Multi-head latent Mamba comprises two modules: input network, multi-head latent interaction (MLI). From the second to the last layers, we adopt a pre-norm operation to prevent overfitting. Specifically, $\mathbf{\tilde{H}}^{(1)} = \mathbf{H}^{(1)}$ and $\mathbf{\tilde{H}}^{(l)} = \mathrm{RMSNorm}(\mathbf{H}^{(l)}), l \in [2,.., L]$. 

\textbf{Input Network}
The input network is designed to map the user context into multiple low-dimensional latent subspaces. This network has two parts, and the first part is a point-wise linear layer with a split operation to obtain the input of downstream Mambas.
\begin{equation}
    \mathbf{X}^{(l)} = \mathrm{Split}(\mathrm{SiLU}(\mathbf{W}^{X} \mathbf{\tilde{H}^{(l)}}))
\end{equation} where $\mathbf{X}^{(l)} = [\mathbf{X}^{(l)}_{1},\mathbf{X}^{(l)}_{2},...,\mathbf{X}^{(l)}_{v}]$ and $\mathbf{X}^{(l)}_{i} \in \mathbb{R}^{n \times d_{c}}$, $\mathbf{W}^{X} \in \mathbb{R}^{vd_{c}\times d}$. $d$ is the dimension of embedding, $d_{c}$ is the dimension of multi-head latent mamba, and $v$ is the head number.

The second part extracts the "item information" for multi-head latent interaction. This part includes a point-wise linear with SiLU activation and the position encoding method RoPE~\cite{10.1016/j.neucom.2023.127063}.
\begin{equation}
    \mathbf{Z}^{(l)} = \mathrm{RoPE}(\mathrm{SiLU}(\mathbf{W}^{Z}\mathbf{\tilde{H}^{(l)}}))
\end{equation} where $\mathbf{W}^{Z} \in \mathbb{R}^{vd_{c}\times d}$, $ \mathbf{Z}^{(l)} \in \mathbb{R}^{n\times vd_{c}}$. According to our experience, incorporating relative position information into the item information only will improve the effectiveness of representation.

\textbf{Multi-head Latent Interaction (MLI)}
The core of the Multi-head Latent Mamba is the multi-head latent interaction. This module utilizes multiple low-dimensional Mamba blocks to extract multi-view historical information and interact it with the item information. We choose multiple low-dimensional Mamba rather than single Mamba with large state size, since the complexity of Mamba scales quadratically with the state size.

\begin{equation}
    \mathbf{Y}^{(l)}_{i} = \mathrm{Mamba}_{i}(\mathbf{X}^{(l)}_{i})
\end{equation}

\begin{equation}
    \mathbf{Y}^{(l)} = \mathrm{Concat}(\mathbf{Y}^{(l)}_{1},...,\mathbf{Y}^{(l)}_{v})
\end{equation} where $\mathbf{Y}^{(l)}_{i} \in \mathbb{R}^{n\times d_{c}}$, and $\mathbf{Y}^{(l)} \in \mathbb{R}^{n\times vd_{c}}$ contains the multi-view historical information of each item in the user context.

As discussed in the Introduction, we try to interact historical and item information to achieve better representation of long and noisy user context sequences. Additionally, the output of Multi-head Latent Mamba needs to combined with a residual contention from $\mathbf{H}^{(l)}$ to facilitate the propagation of low-level features to higher layers throughout the multi-layer architecture. Therefore, a linear projection is required to align the dimensions between interaction results and the input. The formulation of interaction is as follows:
\begin{equation}
    \mathbf{\mathcal{Y}}^{(l)} = \mathbf{W}^{out} (\frac{\mathbf{Y}^{(l)} \odot \mathbf{Z}^{(l)}}{\sqrt{v}}) + \mathbf{H}^{(l)}
\end{equation} where $\mathbf{W}^{out} \in \mathbb{R}^{d \times vd_{c}}$, $\odot$ is element-wise multiplication with a scaling factor $\frac{1}{\sqrt{v}}$. Because we suspect that for large values of $v$, the interaction results grow large in magnitude, pushing the activation function into regions where it has extremely small gradients. To counteract this effect, we scale the multiplication by $\frac{1}{\sqrt{v}}$.

\subsubsection{Feed-Forward Network (FFN)}
We employ a position-wise Gated Linear Units~\cite{shazeer2020gluvariantsimprovetransformer} with SiLU activation as the feed-forward network in the Hydra layer to enhance the modeling of user actions in the hidden dimension. Similarly to Llama~\cite{touvron2023llamaopenefficientfoundation}, bias is not used, and RMSNorm function is employed as pre-norm operation.
\begin{equation}
    \mathbf{\tilde{\mathcal{Y}}^{(l)}} =\mathrm{RMSNorm}(\mathbf{\mathcal{Y}}^{(l)})
\end{equation}
\begin{equation}
    \mathrm{FFN}(\mathbf{\tilde{\mathcal{Y}}^{(l)}}) = (\mathrm{SiLU}(\mathbf{\tilde{\mathcal{Y}}^{(l)}}\mathbf{W}^{gate}) \odot \mathbf{\tilde{\mathcal{Y}}^{(l)}}) \mathbf{W}^{up}) \mathbf{W}^{down}
\end{equation} where $\mathbf{\mathcal{Y}}^{(l)} \in \mathbb{R}^{n \times d}$ is the output of the previous multi-head latent Mamba, $\mathbf{W}^{gate}, \mathbf{W}^{up} \in \mathbb{R}^{d\times d}$, $\mathbf{W}^{down} \in \mathbb{R}^{d\times d}$, and $\odot$ is element-wise multiplication.

Finally, the output of a \emph{l}-th Hydra layer is given by:
\begin{equation}
    \mathbf{H}^{(l+1)} = \mathrm{FFN}(\mathrm{RMSNorm}(\mathbf{\mathcal{Y}}^{(l)})) + \mathbf{\mathcal{Y}}^{(l)}
\end{equation} It should be noted that the primary reason for adopting pre-norm is to facilitate the development of large-scale sequential recommendation models through increased network depth. While post-norm, as used in SASRec, exhibits superior performance in shallow networks, it adversely impacts the learning capacity of deeper networks.

\subsection{Prediction Layer}
Hydra adopts the same prediction layer as SASRec:
\begin{equation}
    \hat{y} = \mathrm{Softmax}(\mathbf{H}_{n}\mathbf{E}^{\top})
\end{equation} where $\mathbf{H}_{n} \in \mathbb{R}^{d}$ is the last hidden representation of the stacked Hydra layers' output sequence, and $\mathbf{E} \in \mathbb{R}^{|\mathcal{I}|\times d}$ denotes the representation of all candidate items. $\hat{y} \in \mathbb{R}^{|\mathcal{I}|}$ represents the probability distribution over the next item in the item set $\mathcal{I}$.

If the embedding layer is used, the item embedding table $\mathbf{E}$ is just the weights of the embedding layer. If a LLM is used as item model, the item embedding table $\mathbf{E}$ needs to be computed by LLM. During inference, we can precompute the item embedding table and store it as an embedding layer to save inference time.
    
\subsection{Multi-domain Extension of Hydra}
\begin{figure}
    \centering
    \includegraphics[width=0.85\linewidth,,height=0.3\textwidth]{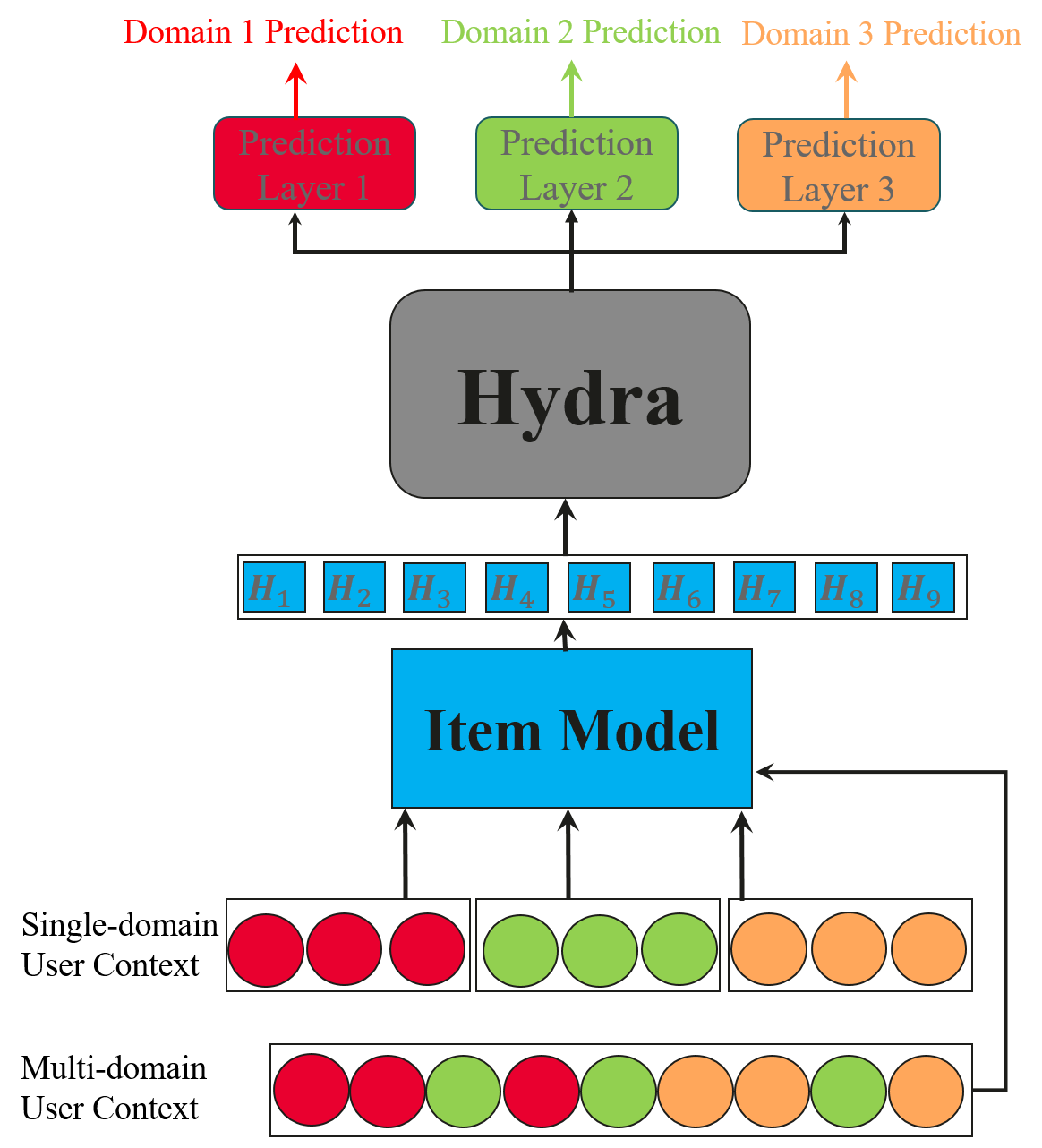}\vspace{-2mm}
    \caption{The architecture of Multi-domain Hydra. Multi-domain Hydra will get two input context sequence, domain-specific and multi-domain user context and predict in each domain. All domains share the stacked Hydra layers.}
    \vspace{-5mm}
    \label{fig:multi-domain}
\end{figure}

The architecture of Hydra multi-domain extension is demonstrated in Figure~\ref{fig:multi-domain}. There are two main differences: \textit{input user context} and \textit{prediction layer}. Based on our experiments, when sufficient computational resources are available, item LLM significantly outperforms embedding layer in enhancing multi-domain recommendation. For more details, please refer to Section~\ref{sec:multi-domain}. 

\subsubsection{Multi-domain User Context}
We largely follow the protocol of $\mathrm{C}^{2}\mathrm{DSR}$~\cite{c2dsr}. Taking two domains as an example, a user has two contexts $U^{1} = \{I^{1}_{1},\dots,I^{1}_{n_{1}}\}$ and $U^{2} = \{I^{2}_{1},...I^{2}_{n_{2}}\}$, which belong to domain 1 and 2, respectively. The multi-domain user context sequence $U = \{I^{1}_{1},I^{2}_{1},\dots, I^{2}_{n_{2}},I^{1}_{n_{1}}\}$ is generated by merging $U^{1}$ and $U^{2}$ in chronological order. Here, $n_{1}$ and $n_{2}$ are the lengths of contexts for domain 1 and domain 2, respectively.

Therefore, if the target item belongs to domain 1, we input $U^{1}$ and $U$ separately to obtain the last hidden representation of the two contexts, \emph{i.e.} $\mathbf{H}^{1}_{n_{1}}$ and $\mathbf{H}_{n_{1}+n_{2}}$. These two vectors are both used in prediction layer.

\subsubsection{Multi-domain Prediction Layer}
Given the observed multi-domain user context sequences, the probability of next item in domain $s$ is defined as:
\begin{equation}
    P(I^{s}_{n_{s}+1} | U) = \mathrm{Softmax}(\mathbf{E}^{s}(\mathbf{H}^{s}_{n_{s}}+\mathbf{H}_{n_{merge}}))
\end{equation} where $I^{s}_{n_{s}+1}$ is an item in domain $s$, and $\mathbf{E}^{s}$ is the item embeddings of all candidate item in domain $s$. The multi-domain training objective will be introduced in Section~\ref{sec:obj}.

\subsection{Complexity Analysis}\label{sec:complex}
Model complexity plays a crucial role in recommendation systems, because online recommendation systems not only demand higher efficiency but also need to process much longer sequence than those encountered in language models. Table~\ref{complexity} demonstrates the superior model efficiency of our proposed Multi-head Latent Interaction. 

In real-world recommendation systems, the length of the user context generally much larger than hidden size, \emph{i.e.} $n >> d$. User context sequences can reach lengths of $10^5$~\cite{zhai2024hstu}. One of the main advantages of Mamba is its ability to address the computational challenges associated with \textit{Transformers} when processing long sequences~\cite{dao2024mamba2}. As shown in Table~\ref{complexity}, the time and space complexity of Mamba and our proposed MLI does not growth with the length of user context, whereas the complexity of Self-Attention scales quadratically with the sequence length. Here, $n$ is the length of context, and $d$ is the dimension of item embedding. While the complexity of Self-Attention is much larger than Mamba and MLI in online recommendation systems. 

Although Mamba-2 can increase state size to improve model capacity, but we find it will significantly increase training and inference time. The complexity of Mamba-2 scales quadratically with the state size. Compared to Mamba-2, MLI can set $vd_{c}^{2} < d^{2}$ to quadratically reduce training and inference cost, such as $v=8, d_{c} = d/4$. Additionally, even a larger number of head can improve model representation capability with less inference cost, because we can take advantage of existing techniques such as \textit{ expert parallelism} to further reduce inference time (from $O(vd_{c}^{2})$ to $O(d_{c}^{2})$).\footnote{\url{https://nvidia.github.io/TensorRT-LLM/advanced/expert-parallelism.html}}

\begin{table}[htp]
\centering
    \caption{Complexity Comparison to Attention, Mamba-2 and Multi-head Latent Interaction. $n$ is context length, $d$ is dimension of embedding, and $d_c$ is dimension of latent spaces.}\vspace{-4mm}\label{complexity}
    \begin{tabular}{cccc}
    \toprule
    &Self-Attention&Mamba-2&MLI\\\hline
    State size&$n$&$d$&$vd_{c}$\\
    Training FLOPs&$n^{2}d$&$nd^{2}$&$nvd_{c}^{2}$\\
    Inference FLOPs&$nd$&$d^{2}$&$vd_{c}^{2}$\\
    Memory&$n^{2}$&$nd$&$nvd_{c}$\\
    \bottomrule
    \end{tabular}
    \vspace{-4mm}
\end{table}

\section{Training for Recommendation Objectives}\label{sec:obj}
The following section provides a detailed description to the training objectives to Hydra, including both single-domain and multi-domain recommendation.

\subsection{Single-domain Objective}
To train Hydra, an objective of generative recommendation is employed~\cite{zhai2024hstu}. Specifically, next item prediction is adopted given the embeddings of the previous items in the context. We utilize the InfoNCE loss~\cite{infonce} during training. 

For any hidden state $\mathbf{H}_{i}$ in the output sequence of the Hydra layers, the positive sample is $\mathbf{H}^{(0)}_{i}$, and the negative samples are randomly sampled from the dataset, excluding the current user sequence. The loss function can be formulated as:
\begin{equation}
    \mathcal{L}_{single} = - \sum_{i=1}^{n} \log \frac{\exp(\mathbf{H}^{(0)}_{i}\mathbf{H}_{i}^\top/\tau)}{\exp(\mathbf{H}^{(0)}_{i}\mathbf{H}_{i}^\top/\tau) +\sum \exp(\mathbf{H}^{'}_{i}\mathbf{H}_{i}^\top/\tau)}
\end{equation} $\mathbf{H}_{i}^{'}$ is the representations of all negative samples for the \emph{i}-th item, and $\tau$ denotes a temperature parameter.


\subsection{Multi-domain Objective}
The objective of multi-domain training consists of two parts: single-domain and cross-domain objectives. We also adopt generative recommendation and InfoNCE loss. The single-domain objective for each domain is identical to that described above.

The loss function of generative recommendation in cross-domain recommendation is similar to single-domain training. The differences lie in the input context and negative sampling strategies. Firstly, the model is trained on the merged multi-domain context sequences. Secondly, negative samples of \emph{i}-th item is sampled from the same domain as \emph{i}-th item. 
\begin{equation}
    \mathcal{L}_{cross} = -\sum_{i=1}^{n_{merge}} \log \frac{\exp(\mathbf{H}^{(0)}_{i}\mathbf{H}_{i}^\top/\tau)}{\exp(\mathbf{H}^{(0)}_{i}\mathbf{H}_{i}^\top/\tau) +\sum \exp(\mathbf{H}^{'}_{i}\mathbf{H}_{i}^\top/\tau)}
\end{equation} where $n_{merge}$ is the length of multi-domain user context, and negative samples representation $\mathbf{H}_{i}^{'}$ is generated by randomly sampling items from the same domain as \emph{i}-th item, excluding the current user sequence.

The total loss function of multi-domain recommendation for Hydra is defined as: 
\begin{equation}
    \mathcal{L}_{total} = \mathcal{L}_{cross} + \sum_{s} \mathcal{L}_{single}^{s}
\end{equation} 

It is worth noting that, generally, the context lengths for multi-domain recommendation are significantly longer than those for single-domain recommendation. This characteristic provides our model with a greater advantage over other attention-based SR approaches, as the computational complexity of our model does not increase with the context length (Section~\ref{sec:complex}).

\section{Experiment}
In this section, we first introduce the basic experimental settings, and then extensive experiments are conducted to address the following research questions:

RQ1: Does the architecture of Hydra itself be able to outperform state-of-the-art sequential recommendation networks?

RQ2: Can Hydra integrate LLMs to improve recommendation performance beyond existing methods that combine LLMs?

RQ3: Is it necessary to fine-tune LLMs for each domain?

After addressing these questions, we investigate the model efficiency of Hydra. Finally, we evaluate Hydra in online scenarios and achieve real-world benefits.


\subsection{Experiment Setups}
\textbf{Implementation Details.} We implement Hydra and all baselines using the code provided by HLLM\footnote{\url{https://github.com/bytedance/HLLM}} and Mamba~\footnote{\url{https://github.com/state-spaces/mamba}}. The expand factor of Mamba is set to 2, and dimension of state is set to 128. The number of negative samples is set to 512. We use the AdamW optimizer with learning rate 1e-3 and weight decay 0.01. The maximum training epoch is 20. We use a linear learning rate warm-up (for the first 5\% steps) followed by cosine annealing. Temperature parameter $\tau$ is set to 0.05 for all experiments. Early stopping strategy is adopted to prevent overfitting with a patience of 3 epochs.

\textbf{Dataset}. 
For offline experiments, we evaluate Hydra on four large-scale datasets from Amazon Reviews 2023~\cite{hou2024bridging}, which is 245.2\% larger than the last version\footnote{\url{https://amazon-reviews-2023.github.io}}: \textit{Books, Movies \& TV, Video Games} and \textit{Toys \& Games}. We follow the preprocessing protocol of HLLM~\cite{chen2024hllmenhancingsequentialrecommendations} and retain only items and the users with at least 5 presences. We only keep item\_id, user\_id and timestamp and ignore other features. The maximum length of context is set to 50. A detailed analysis of these datasets after preprocessing is presented in Table~\ref{tab:data}. 

\begin{table}[htp]
    \centering
    \caption{Statics of Datasets}\vspace{-2mm}
    \begin{tabular}{cccc}
    \toprule
         Dataset&\#User&\#Item&\#Interaction\\\hline
         Books&948,978&966,607&11,544,935\\
         Movies \& TV&691,621&285,252&7,905,714\\
         Video Games&112,130&54,984&973,580\\
         Toys \& Games&519,682&329,920&4,686,250\\
    \bottomrule
    \end{tabular}
    \vspace{-2mm}
    \label{tab:data}
\end{table}

\textbf{Metrics.} We utilize a leave-one-out approach to split the data into training, validation, and testing sets. Performance is measured using Recall@K (R@K) and NDCG@K (N@K). To account for variability, each experiment is repeated 5 times with different random seeds, and we report the average results.

\begin{table*}[htp]
    \centering
    \caption{Evaluation in Public Benchmark Datasets. Recall and NDCG values are averages over 5 random seeds and only keep 2 decimal. The best results on the same dataset are bold with $p<0.05$, and the underline indicates the best result with embedding layer. $\ast$ indicates the result is copied from original paper.}\vspace{-4mm}
    \begin{tabular}{c|c|c|ccccccc}
    \toprule
         Dataset&Item Model&Method&R@10&R@50&R@200&N@10&N@50&N@200&Avg Impv.\\\midrule
         \multirow{9}{*}{Movies \& TV}&\multirow{7}{*}{\textbf{Embedding Layer}}&Mamba4Rec(2024)&3.38&8.68&16.98&1.85&3.00&4.24&+0.0\%\\
         &&HSTU (2024)&7.93&16.45&27.54&4.57&6.42&8.09&+106.38\%\\
         &&HSTU-large&8.59&16.75&27.87&5.01&7.04&8.67&+120.21\%\\
         &&SASRec(2018)&8.55&16.74&27.33&5.14&6.93&8.52&+119.42\%\\
         &&SASRec-1B&8.27&15.90&27.04&4.92&6.77&8.38&+112.71\%\\
         \cmidrule(r){3-10}
         &&Hydra-0.28B&8.62&16.77&27.88&5.16&7.09&8.68&+122.12\%\\
         &&Hydra-0.48B&\underline{8.65}&\underline{16.87}&\underline{27.97}&\underline{5.25}&\underline{7.29}&\underline{8.82}&+124.88\%\\\cmidrule(r){2-10}
         &\multirow{2}{*}{\textbf{Item LLM}}&HLLM-1B(2024)&8.75&16.74&27.96 &4.98&7.24&8.86 &+124.76\%\\
         &&Hydra-0.98B&\textbf{9.00} &\textbf{17.95} &\textbf{28.08} &\textbf{5.97} &\textbf{8.04} &\textbf{9.45} &+142.00\%\\
         \midrule
         \multirow{11}{*}{Books}&\multirow{8}{*}{\textbf{Embedding Layer}}&Mamba4Rec&2.88&7.71&15.56&1.51&2.55&3.72&+0.0\%\\
         &&HSTU-large$\ast$&4.78&10.82&19.08&2.62&3.93&5.17&+49.26\%\\
         &&SASRec&5.35&11.91&21.02&2.98&4.40&5.76&+66.68\%\\
         &&SASRec-1B&5.09&11.11&19.45&2.86&4.17&5.42&+57.41\%\\
         &&HSTU&5.00&11.29&20.13&2.78&4.14&5.47&+57.15\%\\
         &&HSTU-large&5.25&12.03&21.60&2.89&4.36&5.80&+65.90\%\\\cmidrule(r){3-10}
         &&Hydra-0.7B&5.35&12.14&21.78&2.98&4.42&5.84&+68.48\%\\
         &&Hydra-1B&\underline{5.76}&\underline{12.91}&\underline{22.25}&\underline{3.25}&\underline{4.57}&\underline{6.07}&+78.01\%\\\cmidrule(r){2-10}
         &\multirow{3}{*}{\textbf{Item LLM}}&LEARN-7B (2024)$\ast$&4.07 &9.79 &18.74 &2.24 &3.71 &4.83 &+35.40\%\\
         &&HLLM-1B&5.97 &13.61 &23.78 &2.98 &4.64 &6.16 &+15.20\%\\
         &&Hydra-0.98B&\textbf{6.17} &\textbf{13.95} &\textbf{24.58} &\textbf{3.18} &\textbf{5.04} &\textbf{6.75} &+90.47\%\\
    \bottomrule
    \end{tabular}
    \vspace{-4mm}
    \label{tab:main_result}
\end{table*}

\subsection{Evaluation on Public Benchmark Datasets}

\subsubsection{Evaluation with Embedding Layer (RQ1)}
To demonstrate the architecture superiority of Hydra, we first compared it with the baselines using embedding layer on, \textit{Books} and \textit{Movie \& TV}. We follow the traditional sequential recommendation settings as described in the literature~\cite{10.1145/3581783.3611967}. Consequently, an embedding layer with dimension 512 is adopted to encode item IDs to vectors.

The underline in Table~\ref{tab:main_result} indicates the best result with embedding layer. The following conclusions can be drawn from the table:

1) Hydra significantly outperforms existing baselines across all metrics on both datasets when using embedding layer. On the \textit{Movies \& TV} dataset, it surpasses the baseline with only 28\% of the parameters. Meanwhile, on the \textit{Books} dataset, it achieves this with just 70\% of the parameters. 2) With only item id on \textit{Books}, our Hydra significantly outperforms LEARN, which utilizes a frozen 7B LLM to extract text features, and SASRec-1B, which is initialized with pre-trained LLM's parameters. 3)  As our model scales up in parameters, its recommendation performance is further improved. In contrast, SASRec-1B exhibits worse performance when its parameter size increases compared to when it is smaller. 

These results demonstrate the superiority of our model and confirm that the interaction between historical and item information effectively enhances the ability to model user context. Moreover, above results indicate that our model has excellent parameter scalability and holds the potential to become a mainstream architecture for future large-scale recommendation models.


\subsubsection{Evaluation with Item LLM (RQ2)}
To evaluate the ability of Hydra to leverage the "world knowledge" in pre-trained LLMs. We compare Hydra with item LLM to SOTA recommendation models that uses LLMs. Evaluation results are demonstrated in Table~\ref{tab:main_result}. In this experiment, SASRec was initialized with tinyLlama-1T \footnote{\url{https://huggingface.co/TinyLlama/TinyLlama_v1.1}}, and HLLM-1B utilized  two Qwen2-0.5B\footnote{\url{https://huggingface.co/Qwen/Qwen2-0.5B}} to model items and users' interest, respectively. Hydra employed a Qwen2-0.5B to encode flattened text of items into high-dimensional embedding and stacks multiple Hydra layers on top of the pre-trained LLM.

As shown in Table~\ref{tab:main_result}, by fine-tuning an LLM, the performance of Hydra across all metrics has been significantly improved, averagely 7\% on \textit{Movies \& TV} and 7.5\% on \textit{Books}. By integrating a single LLM using the same strategy as HLLM, our approach also significantly surpasses HLLM, achieving an additional average improvement of 6 percentage points. These results indicate that Hydra can more effectively leverage the rich semantic knowledge from existing LLMs and adapt it appropriately into the recommendation task.


\subsection{Multiple Domains Experiment (RQ3)}\label{sec:multi-domain}
We evaluate Hydra in multi-domain recommendation, following the protocol of $\mathrm{C^{2}DSR}$. The maximum length of multi-domain context was set to 200, while that of the single-domain context remained at 50. We selected four popular datasets from the Amazon Reviews dataset. The results are presented in Table~\ref{tab:multi-domain}. "Single" in the table denotes models trained on a single domain only. To save training and validation time, all models had only two layers, and each layer had four heads. Qwen2-0.5B was used as item model. The following conclusions can be drawn from the table.
\begin{enumerate}
    \item Hydra demonstrates superior capabilities for modeling multi-domain behaviors, especially in small domains (\textit{Video Games} and \textit{Toys \& Games}), where the performance gains are particularly significant. This indicates the strong ability of Hydra in knowledge transferring across domains and its effectiveness addressing cold-start problems.
    \item With item LLM, Hydra achieves significant performance improvements in both large and small domains. Meanwhile, some metrics of multi-domain \textit{Hydra + Embedding Layer} on \textit{Books} and \textit{Movies \& TV} datasets are lower than the single-domain models. This highlights the feasibility and necessity of fine-tuning a single LLM across multiple domains, not only substantially reducing training costs but also leading to better multi-domain recommendation performance.
\end{enumerate}

\begin{table*}[htp]
\small
\setlength{\tabcolsep}{1pt}
    \centering
    \caption{Evaluation on Multi-domain Recommendation. The best results are bold. The metrics are Recall and NDCG. "Single" means the model was only trained on relative domain dataset. "Multiple" means training on merged 4 datasets.}\vspace{-4mm}
    \begin{tabular}{c|l|cccc|cccc|cccc|cccc}
    \toprule
          \multirow{2}{*}{\#Domains}&\multirow{2}{*}{Method}&\multicolumn{4}{c}{Books}&\multicolumn{4}{c}{Movies \& TV }&\multicolumn{4}{c}{Video Games}&\multicolumn{4}{c}{Toys \& Games}\\
         &&R@10&R@200&N@10&N@200&R@10&R@200&N@10&N@200&R@10&R@200&N@10&N@200&R@10&R@200&N@10&N@200\\\hline
         \multirow{3}{*}{Single}&SASRec&5.35&21.02&2.98&5.76&8.55&27.33&5.14&8.52&7.74&30.79&4.31&8.42&1.51&9.26&1.91&3.21\\
         &HSTU&5.00&20.13&2.78&5.47&7.93&27.54&4.57&8.09&7.62&31.48&4.15&8.40&2.45&10.89&1.36&2.84\\
         &Hydra + Embedding Layer&5.35&21.78&2.98&5.84&8.62&27.88&5.16&8.68&7.83&31.55&4.36&8.41&3.20&11.59&1.92&3.22\\\hline
         \multirow{3}{*}{Multiple}&HSTU&5.16&20.58&2.94&5.63&7.94&27.55&4.58&8.11&8.43&32.11&5.75&9.08&3.82&12.99&3.03&4.97\\
         &Hydra + Embedding Layer&5.32&21.85&2.97&5.85&8.60&27.85&5.17&8.66&9.78&33.12&6.45&10.03&4.55&13.30&3.63&5.97\\
         &Hydra + Item LLM&\textbf{5.78}&\textbf{22.41}&\textbf{4.41}&\textbf{7.64}&\textbf{9.74}&\textbf{29.04}&\textbf{6.58}&\textbf{10.20}&\textbf{11.66}&\textbf{35.25}&\textbf{7.09}&\textbf{12.35}&\textbf{5.88}&\textbf{15.42}&\textbf{5.32}&\textbf{7.38}\\
    \bottomrule
    \end{tabular}
    \vspace{-2mm}
    \label{tab:multi-domain}
\end{table*}

\subsection{Training Efficiency}\label{exp:efficiency}

\begin{table*}[htp]
\caption{Parameter amount and average one epoch training time in Movies\& TV along with performance. The unit is second and the results are statistic average of all epochs in 5 times training. The best results on the same dataset are bold with $p<0.05$.}\vspace{-4mm}\label{tab:time_param}
\centering
\begin{tabular}{c|c|c|c|cccccc|c}
\toprule
\multirow{2}{*}{Method}&\multirow{2}{*}{\#Param}&\multirow{2}{*}{Emb Size}&\multirow{2}{*}{Avg Time(s)}&\multicolumn{6}{c|}{Metrics}&\multirow{2}{*}{Avg Time Reducing (\%)}  \\
&&&&R@10&R@50&R@200&N@10&N@50&N@200&\\\hline
SASRec&1B&2048&125844.48&8.27&15.90&27.04&4.92&6.77&8.38&0.0\\
HSTU-large&1B&2048&97005.12&8.59&16.75&27.87&5.01&7.04&8.67&-22.9\\\hline
\multirow{3}{*}{Hydra}&0.11B&256&17820.54&8.54&16.65&27.27&5.10&6.75&8.33&-93.7\\
&0.28B&512&12480.67&\textbf{8.65}&\textbf{16.87}&\textbf{27.97}&\textbf{5.25}&\textbf{7.29}&\textbf{8.82}&-90.1\\
&0.63B&1024&30969.54&8.45&16.69&27.48&4.98&6.77&8.39&-75.4\\
\bottomrule
\end{tabular}
\vspace{-4mm}
\end{table*}

Training efficiency is crucial for online recommendation systems. This section aims to demonstrate the superior efficiency of our model by comparing its training time, parameter amount with SOTA baselines. Table~\ref{tab:time_param} presents the result on \textit{Movies\&TV} dataset. We trained all models from scratch five times and averaged the training time across all epochs. In this experiment, our model consist of 16 Hydra layers with 16 heads. All models were trained on 8 GPUs with 32G memory, and we train large models using multi-GPUs with Deepspeed\footnote{\url{https://github.com/microsoft/DeepSpeed}} stage 2 strategy.

The results in Table~\ref{tab:time_param} highlight the exceptional efficiency of our model. With only 11\% of the parameters and less than 7\% of the training time compared to SASRec, our model achieves better average performance. Meanwhile, it outperforms HSTU using only 12.28\% of the training time and 28\% of the parameters. When scaled up to 0.63 billion parameters, our model's training time is only 24.6\% of that of SASRec and 31.92\% of that of HSTU. Additionally, Hydra exhibits much higher convergence speed. We find that SASRec and HSTU can be further improved with longer training, \emph{e.g.} 200 epochs. But Hydra can outperform them within 10 epochs of training. 

The results in Table~\ref{tab:time_param} also demonstrate that the embedding size should be carefully adjusted according to the specific circumstances. Larger embedding size does not necessarily lead to better recommendation performance. For example, when the embedding size is 1024, it not only performs worse but also requires longer training time than the model with an embedding size of 512.

\section{Conclusion}
In this paper, we propose a novel sequential recommendation model, Hydra, designed to enhance sequential recommendations. Hydra models user context by extracting and interacting historical and item information via the proposed multi-head latent Mamba. Our method exhibits exceptional parameter scalability and high efficiency, indicating its potential to become a mainstream architecture for future large-scale recommendation models. Experiments demonstrate that Hydra outperforms conventional sequential recommendation models on several academic datasets with significantly fewer parameters and reduced training time in both single-domain and multi-domain recommendation. To the best of our knowledge, Hydra first demonstrates that fine-tuning a single LLM can enhance recommendation performance across various domains, reducing the cost of fine-tuning for each domain significantly. Real-world online A/B testing further validates Hydra's practical efficiency and applicability, marking a significant advancement in the field of recommendation systems.

\clearpage

\bibliographystyle{ACM-Reference-Format}
\bibliography{sample-base}










\end{document}